\documentclass[12pt,preprint]{aastex}
\usepackage{lineno}
\begin{document}

\parindent=1.0cm

\title{New Blue and Red Variable Stars in NGC 247}

\author{T.J. Davidge\altaffilmark{1,2,3}}

\affil{Dominion Astrophysical Observatory,
\\Herzberg Astronomy \& Astrophysics Research Center,
\\National Research Council of Canada, 5071 West Saanich Road,
\\Victoria, BC Canada V9E 2E7\\tim.davidge@nrc.ca; tdavidge1450@gmail.com}

\altaffiltext{1}{Based on observations obtained at the Gemini Observatory, which
is operated by the Association of Universities for Research in Astronomy, Inc., 
under a cooperative agreement with the NSF on behalf of the Gemini partnership: the 
National Science Foundation (United States), the National Research Council (Canada), 
CONICYT (Chile), Minist\'{e}rio da Ci\^{e}ncia, Tecnologia e Inova\c{c}\~{a}o 
(Brazil) and Ministerio de Ciencia, Tecnolog\'{i}a e Innovaci\'{o}n Productiva 
(Argentina).}

\altaffiltext{2}{This research has made use of the NASA/IPAC Infrared Science 
Archive, which is funded by the National Aeronautics and Space Administration and 
operated by the California Institute of Technology.}

\altaffiltext{3}{This research has made use of the NASA/IPAC Extragalactic Database 
(NED), which is operated by the Jet Propulsion Laboratory, California Institute of 
Technology, under contract with the National Aeronautics and Space Administration.}

\begin{abstract}

	Images recorded with the Gemini Multi-Object Spectrograph (GMOS) on Gemini South 
are combined with archival images from other facilities 
to search for variable stars in the southern portion 
of the nearby disk galaxy NGC 247. Fifteen 
new periodic and non-periodic variables are identified. 
These include three Cepheids with periods $< 25$ days, four 
semi-regular variables, one of which shows light variations similar to those of RCrB 
stars, five variables with intrinsic visible/red brightnesses and colors that are similar 
to those of luminous blue variables (LBVs), and three fainter blue variables, one of 
which may be a non-eclipsing close binary system. The S Doradus 
instability strip defines the upper envelope of a distinct sequence of objects 
on the $(i, g-i)$ color-magnitude diagram (CMD) of NGC 247. 
The frequency of variability with an amplitude $\geq 0.1$ magnitude 
in the part of the CMD that contains LBVs over the 
seven month period when the GMOS images were recorded 
is $\sim 0.2$. The light curve of the B$[e]$ supergiant J004702.18--204739.9, which is  
among the brightest stars in NGC 247, is also examined. Low amplitude variations 
on day-to-day timescales are found, coupled with a systematic trend in mean brightness 
over a six month time interval. 

\end{abstract}

\section{INTRODUCTION}

	Variable stars have long been recognized as important 
probes of stars and stellar systems (e.g. Russell 1912a,b, Hubble 1925). 
Intrinsically luminous periodic variables such as Cepheids, eclipsing binaries (EBs), 
and long period variables are standard candles for distance measurements. 
The study of variable stars in galaxies that sample a range of 
metallicities also has the potential to yield constraints on stellar structure models. 
For example, the properties of massive EBs, such as the fraction 
that are in contact systems, may yield insights into mass loss rates from 
the component stars and the evolution of the close binary systems that play 
important roles in the chemical enrichment histories of galaxies 
(e.g. Sana et al. 2012; Howard et al. 2019). While EBs in the Magellanic Clouds are 
the prime targets for such studies (e.g. Davidge 1988; Almeida et al. 2015) given 
that their component stars have metallicities that are systematically different from 
those in the Galactic disk and are in the nearest extragalactic systems, studies of EBs 
in more distant galaxies will increase the sample size and have the potential to 
identify rare systems (e.g. Prieto et al. 2008). The study of luminous blue 
variables (LBVs) and related objects in galaxies that span a range of 
metallicities will provide insights into the processes that regulate the formation and 
evolution of the most massive stars, and provide constraints on their 
spectrophotometric properties. The environment of LBVs is of particular 
interest for interpreting the nature of these objects (e.g. Smith \& Tombleson 2015; 
Humphreys et al. 2016). 

	NGC 247 is an SAB(s)d galaxy that is in one of the nearest external groups. 
While there are many late-type galaxies within a few Mpc of the Galaxy, NGC 247 is one of 
only a handful that have large-scale structural characteristics that 
hint at tidal interactions within the recent past. The HI disk is truncated 
(Carignan \& Puche 1990), and there are asymmetries in the disk (e.g. Carignan 1985), 
including a lop-sided morphology due to a tidal arm (Davidge 2021). There is also 
evidence for an elevated SFR within the past Gyr (Davidge 2021; Kacharov et al. 2018), 
although the SFR at the present day is relatively low 
(Ferguson et al. 1996), with the peak brightness of the main sequence suggesting 
a marked decline in the past few Myr (Davidge 2006). 

	While we are not aware of published abundance measurements of 
HII regions in NGC 247, it has an M$_K$ that is similar to that of M33 
(Jarrett et al. 2003). The metallicities of young stars in these two galaxies 
should then be similar. The oxygen abundances of HII regions in M33 are 
similar to those in the LMC (e.g. Figure 4 of Toribio San Cipriano et al. 2017), and 
so the metallicity of NGC 247 is likely comparable to that of the LMC. 
There is also evidence that there is not a steep abundance gradient in the 
NGC 247 disk, as the colors of red supergiants (RSGs) 
throughout its disk suggest that the metallicity of these 
objects does not vary by more than $\sim 0.1$ dex over 
most of the young disk (Davidge 2006). Thus, luminous variable stars found throughout the 
disk of NGC 247 should have similar metallicities. 

	The structural characteristics and stellar content of NGC 247 are also of 
interest as it is located within a few hundred kpc of the starburst galaxy 
NGC 253 (e.g. Karachentsev et al. 2003). Along with M82, NGC 253 is one of the 
nearest starburst galaxies, and so is an obvious  
target for understanding the origins and evolution of starburst activity. Unlike 
M82, an unambiguous trigger for starburst activity in NGC 253 has not yet 
been identified. An interaction with a large companion like NGC 247 
is one possibility, and the timing of the onset of elevated SFRs in NGC 247 
and NGC 253 more-or-less coincide (Davidge 2021), even though the SFR of NGC 247 
has declined within the past few million years (Ferguson et al. 1996; Davidge 2006). 

	The variable star content of NGC 247 has not been extensively 
explored, and past studies have targeted only certain types of objects. 
The Aracauria project (Gieren et al. 2005) detected 23 Cepheids in NGC 
247 that span a range of periods (Garcia-Varela et al. 2008). There are hints 
that the distribution of Cepheids throughout the galaxy may not be uniform, as 
only six Cepheids were found in the southern part of NGC 247, while 
Cepheids with periods $< 30$ days were found exclusively in the northern part of 
the disk. The distribution of Cepheids may then track the lop-sided 
morphology of the galaxy. As for other types of variable stars, Solovyeva et al. (2020) 
examined the spectrophotometric properties of two very luminous stars in 
the southern half of NGC 247, classifying one as a candidate LBV 
and the other as a B$[e]$ supergiant. 

	In the current paper we present the results of a search for variable stars 
in the southern disk of NGC 247. The primary dataset consists 
of $g'$ images recorded with the Gemini Multi-Object Spectrograph (GMOS) on Gemini 
South (GS) during a seven month period in 2008 and 2009. The identification 
of OB EBs was assigned a high priority, and the $g'$ filter was selected to 
target stars with early spectral-types. The observing cadence was also set so that 
variability on timescales appropriate for such EBs (i.e. a few days to tens of days) was 
sampled. This sampling is also effective in the detection of other types of 
variables. The GMOS images are supplemented with observations from the 
Canada-France-Hawaii Telescope (CFHT) MegaCam and National Optical Astronomy 
Observatories (NOAO) DECam. 

	The datasets that are used in this paper and the processing that was 
performed on them are the subject of Section 2. The photometric measurements 
and the methodology for identifying objects with variable light measurements 
are discussed in Section 3. Light curves -- phased 
according to periods when possible -- are presented in Section 4, 
while the locations of the variables on color-magnitude diagrams (CMDs) and within 
NGC 247 are examined in Section 5. An investigation of variability among photometrically 
selected early-type stars follows in Section 6. The paper 
closes in Section 7 with a brief discussion and summary of the results.

\section{OBSERVATIONS \& REDUCTION}

\subsection{GMOS Observations}

	NGC 247 was observed with GMOS on GS (Hook et al. 2004) for program 
GS-2008A-Q-205 (PI: Davidge). GMOS is an imaging spectrograph that covers 
an unvignetted $5.5 \times 5.5$ arcmin area on the sky in the 0.36 to $1.1\mu$m 
wavelength range. The GMOS detector has been updated twice since the instrument 
was first commissioned roughly two decades ago. At the time these data 
were recorded the detector was a mosaic of three $2048 \times 4068$ EEV CCDs, 
with a 0.5 mm separation between CCDs. Each pixel subtends 0.073 arcsec on a side.
The pixels were binned $2 \times 2$ during readout for these observations.

	This was a Band 4 program (i.e. a program intended for poor observing 
conditions) that requested as many observations as possible. By virtue of 
its Band 4 status it was executed when the image quality 
was at best fair, and the delivered image quality typically fell 
between 1.0 and 1.5 arcsec full-width at half 
maximum (FWHM), with an upper limit of 1.6 arcsec FWHM. 
The environmental conditions defined for the program specified that observing 
be limited to no more than light cloud cover, and so there are 
modest night-to-night variations in transparency. 
Only one night of observations was deemed unuseable, and this was due 
to extremely high background light levels. Given that the data 
were not always recorded during photometric conditions, all images were calibrated 
to a common photometric reference that was defined by the CFHT MegaCam images 
discussed by Davidge (2006). 

	An observing sequence for each night consisted of a series of 
$g'$ exposures that each had a 300 second integration time. 
The telescope was offset by a few arcsec between 
exposures. A log of the observations is provided in Table 1, where the date of 
observation, the number of exposures recorded per night, 
and the FWHM category (see below) measured from the point spread 
function (PSF) in the final processed image for each night are 
listed. All but one of the images with poorer image quality were recorded during 
July 2008.

\begin{table*}
\begin{center}
\begin{tabular}{lcc}
\tableline\tableline
Date & \# of Exposures & FWHM Category \\
 (UT) & & (arcsec) \\
\tableline
June 28 2008 & 21 & 1.0 \\
June 30 2008 & 16 & 1.5 \\
July 02 2008 & 14 & 1.5 \\
July 07 2008 & 21 & 1.0 \\
July 09 2008 & 11 & 1.0 \\
July 10 2008 & 15 & 1.5 \\
July 11 2008 & 21 & 1.5 \\
July 17 2008 & 6 & 1.5 \\
July 19 2008 & 30 & 1.5 \\
July 25 2008 & 37 & 1.5 \\
July 28 2008 & 9 & 1.5 \\
Sept 01 2008 & 11 & 1.0 \\
Sept 08 2008 & 11 & 1.0 \\
Sept 10 2008 & 22 & 1.0 \\
Oct 07 2008 & 3 & 1.0 \\
Jan 04 2009 & 24 & 1.0 \\
Jan 05 2009 & 6 & 1.0 \\
Jan 06 2009 & 18 & 1.0 \\
\tableline
\end{tabular}
\end{center}
\caption{GMOS Observations}
\end{table*}

	The observations were processed with a standard pipeline 
for CCD images at visible and blue wavelengths. The required calibrations 
were recorded throughout the June 2008 to January 2009 time frame. The two primary 
processing steps were bias subtraction and flat-fielding, with the latter done using 
observations of twilight sky. The bias-subtracted and flat-fielded exposures 
recorded on each night were aligned to correct for dither offsets and then 
averaged together. Any exposures that had obvious departures in image quality and/or 
transparency for that night were not included when constructing nightly means.

\subsection{Archival Observations}

	Observations from other facilities are used to expand the time coverage and 
extract broad-band colors. Extending the time coverage is of particular interest among 
variable stars with light variations that occur over multi-year time scales and/or in 
an erratic manner. Colors are also of obvious interest to characterize the 
variables. Relevant properties of the archival datasets are summarized 
in Table 2. A brief dicussion of each supplemental dataset is provided in the following 
sub-sections.

\begin{table*}[!ht]
\begin{center}
\begin{tabular}{lccl}
\tableline\tableline
Instrument & Filters & Dates & Program ID \\
 & & (UT) & \\
\tableline
CFHT MegaCam & $g', r', i'$ & 12/23/2003 & 03BC03 \\
 & $g'$ & 11/20/2003 & \\
 & & & \\
CTIO DECam & $g$\tablenotemark{a} & 11/06/2015 & 2012B-0001 \\
 & & & \\ 
SPITZER IRAC & [3.6],[4.5] & 03/18/2014 & 57359 \\
 & & 10/10/2014 & \\
 & & & \\
\tableline
\end{tabular}
\end{center}
\caption{Archival Data}
\tablenotetext{a}{Observations in other filters were obtained with DECam, but these are 
not considered in this paper as they were recorded during non-photometric conditions.}
\end{table*}

\subsubsection{MegaCam}

	Images of NGC 247 in $g'$, $r'$, and $i'$ were recorded with the CFHT 
MegaCam (Boulade et al. 2003) on December 23, 2003. These data were discussed previously 
by Davidge (2006), and were recorded $\sim 3$ years before the 
GMOS observations, thereby allowing long term trends in brightness to be 
examined. These images were also recorded during photometric conditions and have an 
angular resolution that is comparable to that of the GMOS images. Therefore, they 
are well-suited for setting the photometric calibration of the GMOS data. 
The use of a common photometric reference also means that the identification 
of variables is done in a pseudo-differential -- as opposed to absolute -- manner. 
As the MegaCam images of NGC 247 were all recorded on the same night then they 
can be used to measure colors at a single epoch that are free of phase-related variations.

	MegaCam $g'$ observations of NGC 247 were also recorded 
on November 20, 2003. These were not considered by Davidge (2006) as they did not 
pass the seeing requirements specified in the original proposal. Still, these data 
can be used to obtain photometry of the variables, and further expand the time coverage.
Both sets of MegaCam images were processed with the ELIXIR pipeline (Magnier \& 
Cuillandre 2004).

\subsubsection{DECam}

	Images of NGC 247 were recorded in $g$ 
with the CTIO DECam (Flaugher et al. 2015) in 2015 as part of program 2012B-0001 
(PI: Frieman). The DECam images were recorded $\sim 7$ years after the GMOS observations, 
and so provide important constraints on long-term variability. Processed version 
of the DECam images were downloaded from the NOIRLab archive. 

	When compared with the other observations used in this paper, 
the DECam data have a short exposure time (90 seconds in $g$). These data were also 
recorded during non-photometric conditions when the seeing was poor, with 
FWHM $\sim 2$ arcsec. The photometric calibration of the DECam $g$ data 
was set using the same measurements obtained from the CFHT MegaCam observations 
that were used to calibrate the GMOS observations. 
Observations of NGC 247 were also recorded with DECam through other filters. However, 
given the indifferent image quality and the non-photometric conditions, the observations 
recorded with DECam through those filters were not considered. 

\subsubsection{IRAC}

	[3.6] and [4.5] images of NGC 247 were obtained with the 
SPITZER (Werner et al. 2004) IRAC (Fazio et al. 2004) as part of the SPIRITS survey 
(Kasliwal et al. 2017). Observations at these wavelengths are of interest to characterize 
the IR SEDs of variables and identify objects that might have warm circumstellar 
dust shells. The images from the SPIRITS survey sample a number of epochs, and 
the two datasets recorded in 2014 with astronomical observation request identifiers 
50548736 and 50548992 that were discussed by Davidge (2021) were adopted for the present 
work. The processing of these data was discussed by Davidge (2021).

\section{PHOTOMETRIC MEASUREMENTS AND VARIABLE STAR IDENTIFICATION}

\subsection{Photometry}

	The distribution of FWHMs among the nightly images 
is approximately bimodal, with peaks near 1 arcsec and 1.5 arcsec. That the FWHMs 
typically equal or exceed 1 arcsec is due to the relaxed nature of the requested 
observing conditions, which are a defining characteristic of Gemini Band 4 programs. 
Given the bimodal distribution of FWHM values, the images were divided into two groups. 
One group (images with FWHM $\sim 1$ arcsec) constitute the core dataset for the 
identification of variable objects based on the dispersion in their magnitude 
measurements, while the second group (images with FWHM 
$\sim 1.5$ arcsec) provide supplementary photometric measurements 
of variable stars that were identified in the first group. 
The FWHM classification of the images recorded on each night is listed in Table 1. All 
of the images in the 1.5 arcsec FWHM group were recorded in July 2008. 

	A median image of all frames in the 1 arcsec FWHM group was constructed 
after adjusting for frame-to-frame differences in sky transparency, and the 
result served as a reference for photometric measurements and 
the assessment of variability. As it is constructed 
from ten frames, the statistical noise throughout much of the median image is lower 
than in individual frames. However, source confusion is by far the dominant 
obstacle for detecting stars in dense parts of the NGC 247 disk.

	Stellar brightnesses were measured with the PSF fitting 
program ALLSTAR (Stetson 1994). Tasks in DAOPHOT (Stetson 1987) were used to 
identify stars, estimate preliminary magnitudes, and construct 
the PSF from a sample of bright, isolated stars. The photometric 
calibration is based on the brightnesses of isolated stars 
with magnitudes measureed from the CFHT MegaCam $g'$ image.

	Unresolved stars in NGC 247 form a complex background that 
complicates stellar photometry, and light from this background was removed 
using an iterative process. First, an initial set of photometric measurements 
was obtained with the background light in place. The photometered stars were 
subtracted from the initial image, and the result was smoothed with a median filter to 
suppress artifacts of the subtraction process. The resulting smoothed image tracks the 
light from unresolved stars, and was subtracted from the initial image 
to produce an image in which unresolved light has been removed. Final photometric 
measurements were then made from the background-subtracted image.

	The source catalogue obtained from the median image was adopted for 
photometry on each of the nightly images. A potential shortcoming of 
the use of such a master source catalogue is that eruptive variables 
such as novae and background supernovae that fall above the 
detection limit on only some nights will likely be missed. 
However, a benefit is that all of the variables found should 
be recoverable in images recorded at other epochs, and this is borne out 
in the high recovery rate of variables in the MegaCam and DECam images (see below) 
that were recorded years before/after the GMOS observations. 

	The brightnesses of sources in the MegaCam and DECam images were also measured 
with ALLSTAR. As the depths and angular resolutions of the MegaCam and DECam images 
differ from those in the GMOS observations, separate source catalogues were generated 
for each of these datasets. The brightnesses of variables that were identified from the 
GMOS observations were then extracted from the MegaCam and DECam photometric catalogues. 

\subsection{Variable Star Identification Using Dispersion in Photometric Measurements}

	The dispersion in magnitudes measured over a range of epochs is an obvious 
criterion for identifying variable stars. In the current study, sources that had a 
dispersion in night-to-night measurements that departed from the magnitude measured 
in the 1 arcsec FWHM median image at the 5$\sigma$ or higher significance level were 
tagged as candidate variables. The threshold dispersion was set by calculating 
the night-to-night scatter in the brightnesses of stars having similar 
magnitudes, but with $\sigma -$clipping applied to reject variable stars. 
There are numerous other criteria for selecting variables, and 
in Section 6 the sample of variable stars is expanded further by applying 
slightly relaxed detection criteria in a region of the CMD associated with LBVs.

	While dispersion in the photometry is the 
primary criterion for identification as a variable star in this section, 
other constraints were also imposed. Candidate variables had to 
be detected in all ten of the 1 arcsec FWHM images. In 
addition, the images were inspected to identify objects where the photometry might 
be skewed by diffraction spikes from very bright stars, as well as transient 
events such as cosmic rays or trails from minor planets and artificial satellites. 
Finally, objects that were located in or close to the gaps between CCDs or 
were close to the detector edges were removed from the sample.

	While a 5$\sigma$ criterion that also requires detection 
in all ten 1 arcsec FWHM images sets a high threshold for selection, 
it ensures (1) that the light variations in the final sample are not the result of 
statistical flucuations in the photometric measurements, and (2) that a light 
curve could be constructed with at least ten points. The strict identification 
criteria also facilitate the detection of variables in the archival datasets, and 
all of the variables discovered in this study were recovered in at least one of 
the two MegaCam $g$ images. Two of the brightest variables were saturated in 
the December 2003 $g$ image, and so were not recovered with those data. 
All but one of the faintest variables was recovered from the DECam $g$ observations. 

	The photometry of the variables that were identified from the 1 arcsec 
images was supplemented with photometric measurements made from the 1.5 arcsec 
FWHM images. The faint limits of the 1.5 arcsec FWHM images tend to be a few tenths of a 
magnitude brighter than that of the 1 arcsec FWHM images, and sources that are close to 
the faint limit of the 1 arcsec images when they are near the low point of their 
light variations were not recovered in some of the 
1.5 arcsec FWHM images. Still, the majority of variables found here 
are bright enough that measurements made from the 1.5 arcsec images tend not to 
contribute additional scatter to the light curves.

	The co-ordinates and photometric properties of the variables identified using 
the procedures described above are listed in Table 3. The co-ordinates were obtained 
from the world co-ordinate system information in the GMOS image headers. 
The co-ordinates measured from bright stars in the GMOS image are consistent 
with those in the 2MASS Large Galaxy Atlas (Jarrett et al. 2003) image of NGC 247  
to within 1 -- 2 arcsec, suggesting that the co-ordinates 
in Table 3 are reliable to $\pm 2$ arcsec. The magnitudes listed in Table 3 are 
from the 1 arcsec FWHM median image, and so are 'typical' brightnesses 
during the second half of 2008. The periods and light curve 
types listed in Table 3 are discussed in the next section. Finding charts that show the 
$15 \times 15$ arcsec area centered on each variable are shown in Figure 1. 

\begin{table*}[!ht]
\begin{center}
\begin{tabular}{rcccccc}
\tableline\tableline
ID & RA & Dec & $g'$\tablenotemark{a} & Period & Type\tablenotemark{b} & Other \\
 & (2000) & (2000) & & (days) & & Names \\
\tableline
Var001 & 00:47:15.4 & --20:47:35 & 20.2 & & cLBV & \\
Var002 & 00:47:00.7 & --20:46:39 & 21.5 & & SRb & \\
Var003 & 00:47:05.8 & --20:48:05 & 21.5 & & cLBV & \\
Var004 & 00:47:08.8 & --20:50:56 & 21.1 & & BV & \\
Var005 & 00:47:10.5 & --20:48:50 & 21.6 & 13.0? & cLBV & \\
Var006 & 00:47:12.7 & --20:50:00 & 21.7 & 33.3? & CBS? & \\
Var007 & 00:47:08.7 & --20:47:16 & 21.8 & & SRb & \\
Var008 & 00:47:10.3 & --20:51:14 & 22.0 & 73.3 & Cep & cep022 \\
Var009 & 00:47:11.3 & --20:49:01 & 22.5 & & cLBV & \\
Var010 & 00:47:11.5 & --20:48:47 & 22.7 & 33.2 & Cep & cep011 \\
Var011 & 00:47:04.8 & --20:49:45 & 23.6 & & SRb & \\
Var012 & 00:47:09.0 & --20:48:24 & 23.2 & & SRd & \\
Var013 & 00:47:10.5 & --20:50:20 & 22.9 & 22.9 & Cep & \\
Var014 & 00:47:16.3 & --20:50:26 & 22.7 & 22.5 & Cep & \\
Var015 & 00:47:17.7 & --20:47:48 & 23.1 & 20.3 & Cep & \\
\tableline
\end{tabular}
\end{center}
\caption{Variables Detected Using the Dispersion in the Photometric Measurements}
\tablenotetext{a}{Magnitude measured from the median 1 arcsec FWHM $g'$ GMOS image.}
\tablenotetext{b}{cLBV = candidate luminous blue variable; SR = semi-regular variable; 
Cep = Cepheid; CBS = close binary system.}
\end{table*}

\begin{figure}
\figurenum{1}
\epsscale{1.0}
\plotone{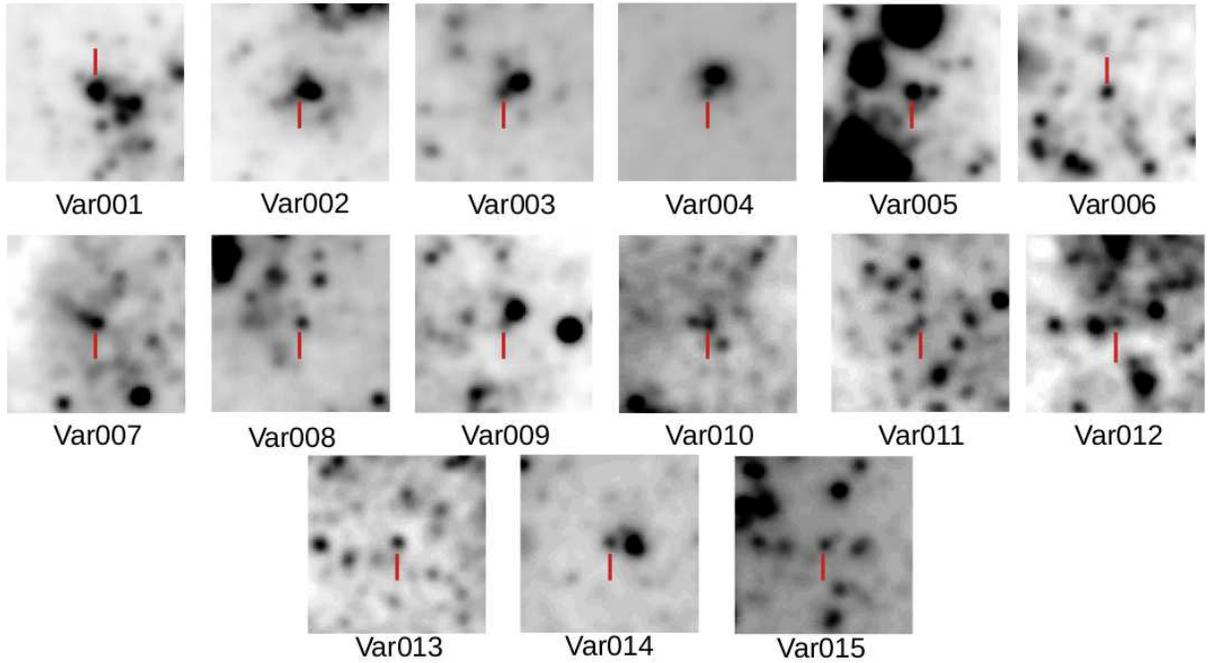}
\caption{Finding charts for the variables listed in Table 3, which were identified 
using the $5\sigma$ dispersion criterion discussed in Section 3. 
Each image is $15 \times 15$ arcsec on a side, with 
North at the top and East to the right. The variable stars are centered in 
each frame, and are at the end of the red line. These images were 
extracted from the median $g'$ 1 arcsec FWHM image.}
\end{figure}

	The area observed with GMOS contains a modest number of 
previously identified variable stars. Two Cepheids (cep011 and cep022) 
discovered by Garcia-Varela et al. (2008) were recovered 
using the procedure described above. There is an 
offset with respect to the Garcia-Varela et al. (2008) co-ordinates of 2 arcsec in 
declination, and 1 sec ($\sim 14$ arcsec) in right ascension. 

	Garcia-Varela et al. (2008) found four other Cepheids in the area imaged 
with GMOS, and these were not recovered independently with the GMOS observations. 
One of these (cep013) was missed because of its proximity to the edge of the 
GMOS science field. An examination of the GMOS photometry of this object reveals 
variations in its brightness that exceed the 5$\sigma$ 
detection criterion as well as a light curve that is consistent with Cepheid 
variability. Thus, it would have been detected if it were further from the 
edge of the GMOS science field. Lacking finder charts, we have not been able to 
identify unambiguously the other three previously-detected variables (cep008, 
cep016, and cep018) in the median $g'$ image, and so the brightness variations 
in the GMOS data can not be examined to assess why these stars 
evaded detection. We suspect that the failure to independently recover 
these variables is due to the shorter time coverage of the GMOS data 
(7 months for GMOS versus up to 2.3 years for the data used by Garcia-Varela et al. 
2008), coupled with stochastic effects when sampling their light curves.

	Four of the stars in Table 3 are classified as candidate LBVs (cLBVs), and 
this classification should be regarded only as a place-holder. 
{\it Bona fide} LBVs form a diverse group of objects 
(reviews by Humphreys \& Davidson 1994 and Smith 2014), and are just one type 
of intrinsically bright blue variable star. Humphreys \& Davidson (1994) 
review the photometric properties of LBVs, and identify four distinct time scales 
for variations, with the largest amplitude variations occuring over time scales of 
decades or centuries. Barring a fluke detection of such a multi-magnitude 
eruption from an LBV, the observations discussed here will more likely 
detect the variations on the order of a few tenths of a magnitude 
that are seen over timescales of a few months to years in the light curves of some LBVs. 
Martin \& Humphreys (2017) examine the photometric properties of LBVs and candidate 
LBVs in M33 over a five year baseline and find 0.1 - 0.8 magnitude spreads in $V$. Other 
types of intrinsically bright blue variables show a similar range in $V$ magnitudes.

	Solovyeva et al. (2020) examined spectra and photometry of two luminous 
early-type stars, and both are in the area of NGC 247 that was observed with GMOS. One 
of these (J004703.27-204708.4) was not detected as a variable in the present study 
using the $5\sigma$ criterion described in this section, although 
evidence is presented for moderate variability in the light curve of this star 
in Section 6. The other star fell in the gap between CCDs in the 
GMOS detector, and so has uncertain photometry.

\section{LIGHT CURVES, PERIODS, AND PHASING}

	The nightly GMOS magnitude measurements 
of the sources listed in Table 3 are shown in Figure 2. The random 
uncertainties in the individual measurements are typically 
a few hundredths of a magnitude, and it should be recalled (Section 3.2) 
that a criterion for selecting objects as variables was that the dispersion in 
the various measurements exceed this at the $5\sigma$ level.
The dispersion in the measurements that allowed these objects to be flagged as variables is 
thus clearly evident. These light curves were used to sort the sample into two groups: 
(1) obvious periodic variables, and (2) objects that are either non-periodic or 
that have a period that can not be determined from the existing data. 
The objects in each group are discussed in the following sub-sections.

\begin{figure}
\figurenum{2}
\epsscale{1.0}
\plotone{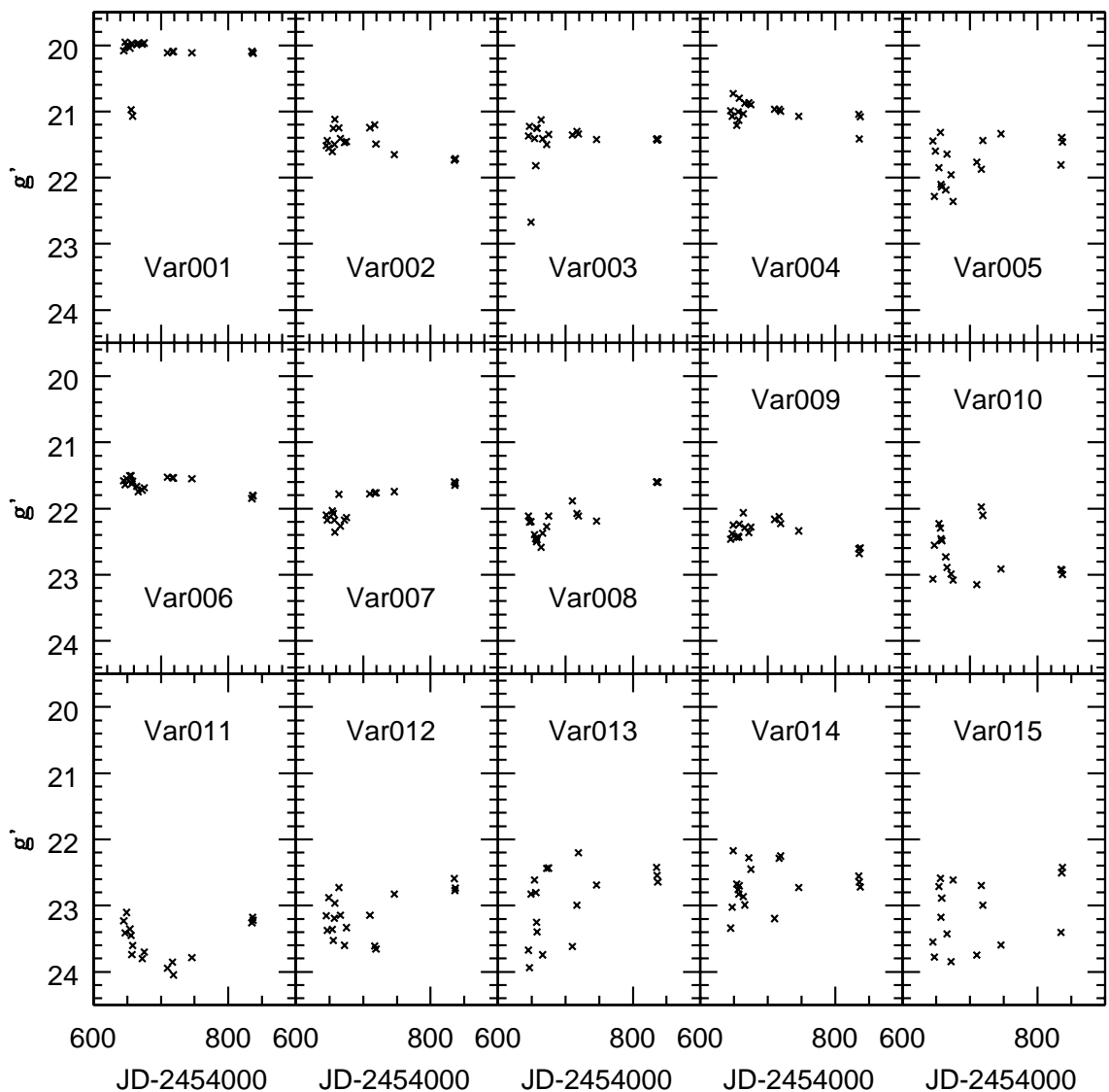}
\caption{Unphased light curves of the variable sources that were identified using the 
dispersion criterion in Section 3. Only measurements made from the GMOS images are shown 
to highlight the variations in brightness that served as the 
basis for their identification as {\it bona fide} variables. 
The random uncertainties in individual measurements is typically 
a few hundredths of a magnitude, and so is -- by definition 
(Section 3.2) -- much smaller than the amplitude of the light 
variations. Some of the light curves show signs of periodic behaviour, while 
others show no evidence of periodicity in the time interval sampled with GMOS.}
\end{figure}

\subsection{Periodic and Possibly Periodic Variables}

	The periods of some stars can be determined 
directly from the measurements in Figure 2. The phased light curves of 
these stars, constructed with the periods listed in Table 3, are 
shown in Figure 3. As these are a disparate group of objects then zero phase coincides 
with the first date that images were recorded with GMOS (June 28, 2008). 

\begin{figure}
\figurenum{3}
\epsscale{1.0}
\plotone{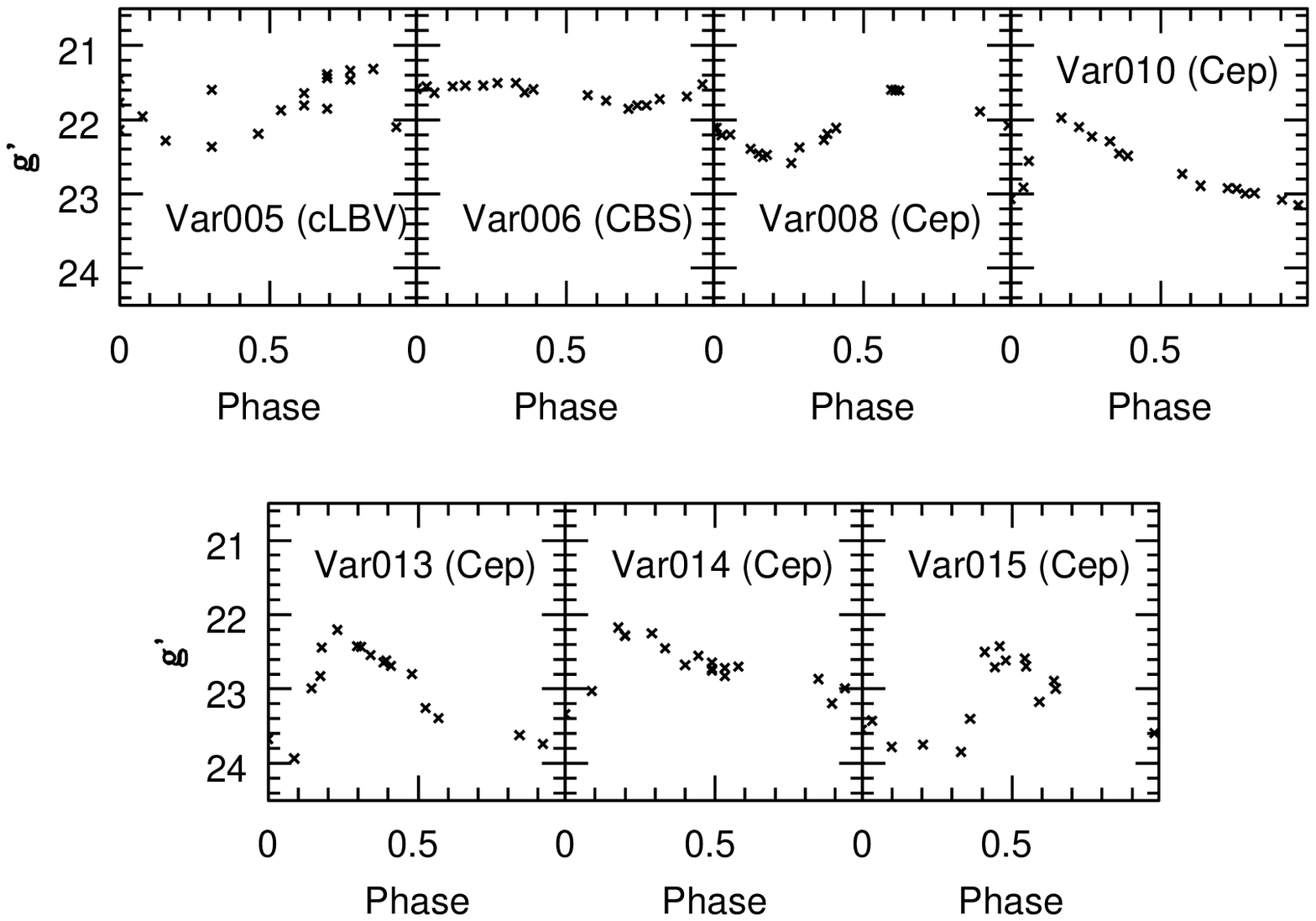}
\caption{Phased light curves. Given the disparate nature 
of the variables, phase zero coincides with June 28, 2008 (i.e. the first date 
of observation with GMOS). The light curves of Var008, Var014, Var013, Var010, and Var015 
are indicative of classical Cepheids. Based in large part on 
its location in the CMD we suggest that Var006 is a CBS. Var005 is included in this 
figure given the periodic nature of its light variations in Figure 2. However, the 
scatter in the light curve indicates that the period estimated for Var005 is highly 
uncertain, and this star is classified as a candidate LBV.}
\end{figure}

	The phased light curves of five of the stars (Var008, Var010, Var013, Var014, 
Var015) coupled with their broad-band colors (Section 5) suggest that they are classical 
Cepheids. The two brightest Cepheids identified from the GMOS images were also identified 
by Garcia-Varela et al. (2008), and these are stars cep011 and cep022 in their Table 5. 
The three faintest Cepheids in our sample were not detected by Garcia-Valera 
et al. (2008). These likely evaded detection in that study because 
they are faint and in moderately crowded environments. Cepheids 
with similar periods were discovered by Garcia-Valera et al. (2008), but in the northern 
parts of NGC 247, where the stellar density tends to be lower than in the south. 

	The light curve of Var006 in Figure 2 shows periodic low amplitude variations. 
While the phased light curve of Var006 is reminiscent of that 
of a Cepheid with a broadened maximum, it is likely not a Cepheid. 
It is near the bright end of stars in our sample, 
and the 33.3 day period is shorter than might be expected 
for an intrinsically bright Cepheid, although the difference in magnitude between 
Var006 and Var008 is within the $\pm 0.5$ mag scatter (Storm et al. 2011) in the 
period-luminosity relation. The light curve of this star also has a modest amplitude 
when compared with that of other Cepheids. Most importantly, in Section 5 it is 
shown that its location in the CMD places it well blueward of the Cepheid 
instability strip, making it an object with an early spectral type. 

	While not shown in Figure 2, the MegaCam and DECam 
observations indicate magnitudes that are consistent with those made with GMOS, and 
so there is no evidence for large amplitude variations over timescales of many years. 
This leads us to conclude that while Var006 is a bright early-type variable, it 
is likely not a LBV as it is also well below the S Dor instability strip (Section 5).
Given the low amplitude, periodic nature of the light 
variations, we suggest that Var006 is a close binary system (CBS), with the light 
variations due to temperature variations on one star in a contact or near-contact 
binary system. The inclination of the system is such that eclipses do not occur. A 
massive CBS interpretation is consistent with a system of two stars 
with M$_V \sim -5.5$, and a spectropscopic search for signatures of two stars 
would confirm or negate this possibility. If it is a massive system 
undergoing mass transfer then spectra might also reveal signatures of 
system-wide mass loss due to mass flow through the L$_2$ Lagrangian point 
(e.g. Webbink 1976).

	Despite showing hints of periodic variations in Figure 2, 
the phased light curve of Var005 in Figure 3 suggests that 
these may only be short-lived quasi-periodic variations. There is considerable 
scatter in the phased light curve of this star, indicating that the 13 day period 
estimate is uncertain, even though it appears to be moderately well constrained 
by the July 2008 observations. Var005 is one of the brightest variables 
identified here, and has colors that are indicative of an early-type star.
The location of Var005 on the CMD constructed from the MegaCam observations 
is $\sim 1$ magnitude below the S Dor instability strip (Section 5). However, 
it is $\sim 1.5$ magnitudes brighter in the GMOS observations, placing it in the 
region occupied by LBVs. Hence, we suggest that it is a candidate LBV that had 
quasi-periodic variations in June and July 2008.

\subsection{Semi-regular and Non-periodic Variables}

	Light curves of the sources that do not show evidence 
of clear periodicity in the GMOS observations are shown in Figure 4. DECam and MegaCam 
measurements are also shown so that trends over decadel timescales can be examined. 
Based on these light curves, combined with locations in the CMD (Section 5), 
these are identified as probable semi-regular red and yellow supergiant variables, 
while others are potential LBVs. 

\begin{figure}
\figurenum{4}
\epsscale{1.0}
\plotone{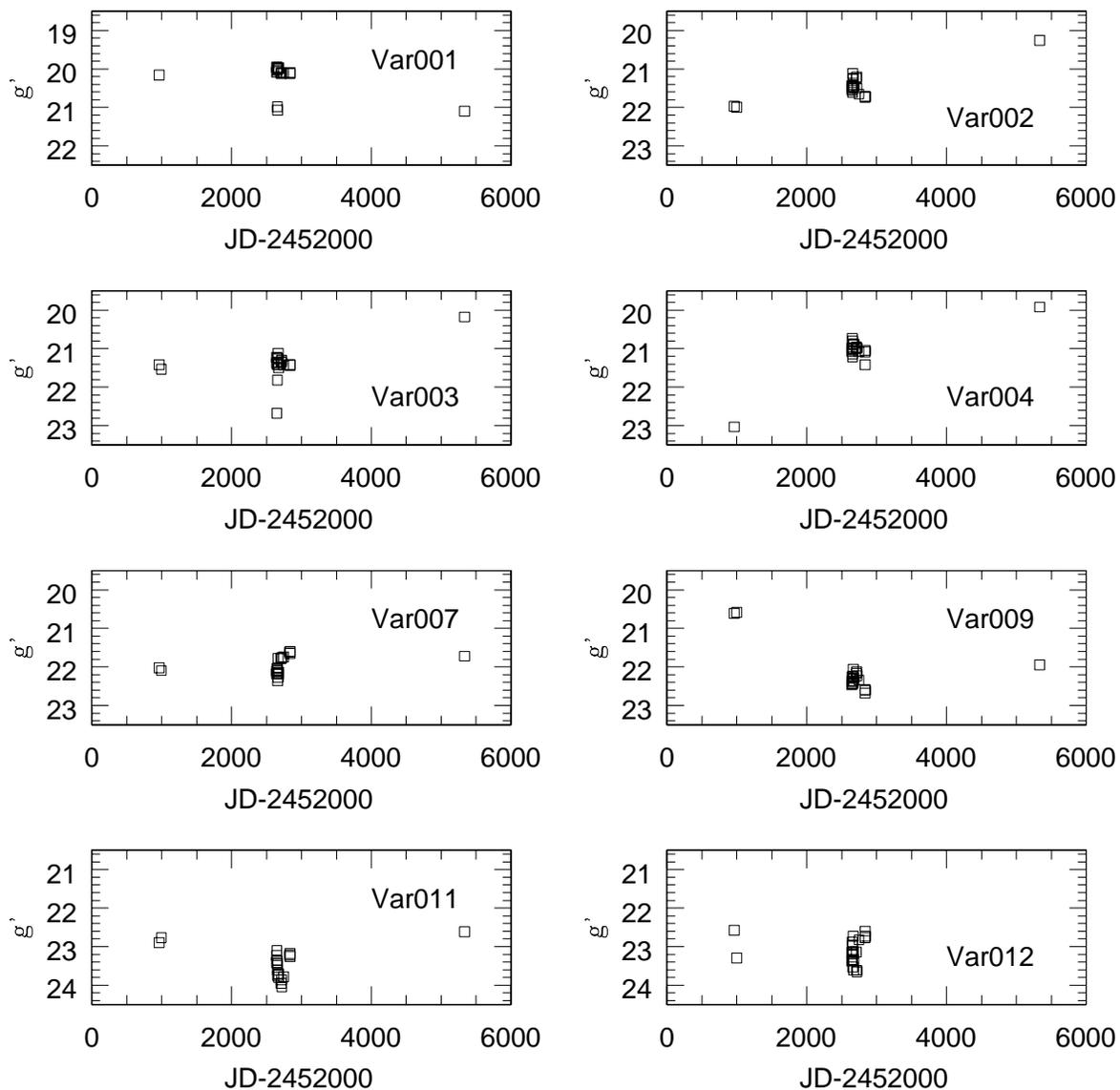}
\caption{Unphased light curves of variables without periods. Photometry 
from the 2003 MegaCam observations and the 2015 DECam observations are also shown. 
The scatter in the GMOS measurements that served as the basis for the identification 
of these objects as variables is clearly evident. The inclusion of the MegaCam and 
DECam photometry reveals trends in visible light that extend over decade-long timescales 
in the light curves of Var002, Var003, Var004, and Var009.}
\end{figure}

	There are episodic variations in the GMOS light curve of Var002 in Figure 2, with 
considerable night-to-night scatter. The amplitude of the light 
variation increases with the inclusion of photometry from the MegaCam and 
DECam images, and there is a clear long-term trend in brightness in Figure 4. 
Var002 has one of the reddest $g'-i'$ colors (next section). The photometric 
characteristics are consistent with light variations from supergiants that show 
semi-regular light variations. We thus classify Var002 SRb.

	The light curve of Var011 is unique among the variables found 
in this paper. There is a 0.8 magnitude decline in light level over a time span of 
tens of days in the GMOS observations, although at the end of the $\sim 200$ 
day GMOS observation sequence the light level had recovered to its initial 
value. The MegaCam and DECam brightnesses of this 
star in Figure 4 indicate that the initial and final light levels in the GMOS 
data are `typical' brightnesses for this star.

	Light variations of this nature and duration are reminiscent of those 
in R CrB stars. The dominant features in the light curves of 
classic R CrB variables are multi-magnitude dips in light levels over long time 
spans. These are accompanied by frequent dips in brightness that amount to 
a few tenths of a magnitude. The large scale variations in light levels are 
thought to be due to extinction by dust, while the more modest variations may be 
due to pulsations (e.g. Clayton 1996). R CrB stars are H-deficient 
supergiants with a typical intrinsic magnitude M$_V \sim -5.5$, and so 
such a star in NGC 247 should have $g \sim 22.5$, which is 
consistent with the measured brightness of Var011 outside of the light dip seen here.
A potential problem with the R CrB interpretation for Var011 is that these stars 
have excess IR emission from circumstellar dust (e.g. Lambert et al. 2001). The 
[3.6]--[4.5] color of Var011 does not show signs of an IR excess (Section 5), 
suggesting that if hot dust is present then it does not dominate the signal at these 
wavelengths. The nature of the variability observed in Var011 is thus not clear. Spectra 
would reveal if this star is deficient in Hydrogen.

	The light variations of Var012 in Figure 3 are suggestive of 
periodic activity, although a period could 
not be found with these data. If the observations are restricted to those 
recorded in June -- July 2008 then a Cepheid-like light curve with a 14 -- 17 day 
period is found, with preference for longer periods. 
However, when the entirety of the 7 month GMOS dataset is 
considered the construction of a well-defined phased light curve is elusive. 
There is a $\sim 0.6$ magnitude difference in the November and December 2003 MegaCam 
observations in Figure 4, and this is comparable to the dispersion 
in the GMOS observations, hinting that the amplitude seen 
in the GMOS observations is at least stable, if not periodic. 
Based on these light variations we suggest that Var012 is a semi-regular variable. The 
location on the CMD places it near the blue edge of the Cepheid instability strip 
(Section 5), leading to a proposed classification of SRd.

	Multi-year trends in brightness are seen in the light curves of 
Var001, Var003, Var004, and Var009 in Figure 4. All have blue $g'-i'$ colors, and are 
intrinsically bright sources at visible/red wavelengths. The light 
curves of Var001 and Var003 show $\sim 1$ magnitude drops at two different epochs 
in the GMOS data, while Var009 shows a slower paced drop in its light level in Figure 2. 
Var004 shows a steady increase in brightness throughout the 12 year time span bracketed 
by the MegaCam and DECam images.

	The blue variables tend to be in complex 
environments with bright nearby neighbors, and so their photometry is prone to 
night-to-night seeing variations. However, the dips in the light 
levels in Var003 and Var001 occur on two different nights 
that had different image qualities. This suggests that these dips 
are not related to seeing variations, which could affect photometry in 
crowded areas. The blue variables have mean intrinsic brightnesses M$_V \sim -9$ 
(Var001), $-8$ (Var003), $-7$ (Var004), and $-5.5$ (Var009) if they are at the 
distance of NGC 247, and this is consistent with them being 
massive evolved stars. Based on the location of these objects in the CMD, 
we classify Var001 and Var003 as {\it potential} LBVs in Table 3. Var004 and 
Var009 are classified as blue variables - these object are likely evolved massive stars 
that are not LBVs. 

\section{PHOTOMETRIC PROPERTIES: LOCATIONS IN CMDS AND ON THE SKY}

	The archival observations described in Section 2 span a 
broad range of wavelengths, and can be used to place 
the variable stars on a CMD, albeit with photometry that samples 
only one epoch. Not only do photometric measurements that span a 
range of wavelengths provide clues into the nature of the variables, but they can also 
be used to examine their immediate environment; for example, the detection of 
IR excess radiation could be suggestive of mass deposition into a circumstellar envelope. 
Further insights into the nature of the variables, including their evolutionary 
state and physical characteristics, can be gleaned by examining 
their locations in NGC 247, such as proximity to star-forming regions. 

	Photometric measurements of the variables 
taken from the MegaCam and IRAC images are compiled 
in Table 4. The majority of measurements at visible/red wavelengths are 
from the December 2003 MegaCam images and give colors only at that epoch. The SEDs 
of variables will likely change with phase, and the MegaCam and IRAC images 
do not sample the same phase. The $1\sigma$ random uncertainties as estimated by 
ALLSTAR are shown in brackets beneath the measurements. Uncertainties in the 
absolute photometric calibration of the MegaCam measurements are on the order of a 
few hundredths of a magnitude. The intrinsic angular resolution of the SPITZER images is 
poorer than that of the MegaCam observations. The potential for blending, especially 
in compact star forming regions that contain the candidate LBVs, thus should not be 
discounted. Five variables were not recovered in the SPITZER images. 

\begin{table*}[!ht]
\begin{center}
\begin{tabular}{cccccc}
\tableline\tableline
ID & g'\tablenotemark{a} & g'---r'\tablenotemark{a} & r'--i'\tablenotemark{a} & [3.6]--[4.5]\tablenotemark{b} & $g'$--[3.6] \\
\tableline
Var001 & 20.192 & --0.064 & 0.075 & --0.02 & 4.04 \\
 & ($\pm 0.02$) & ($\pm 0.03$) & ($\pm 0.03$) & ($\pm 0.07$) & ($\pm 0.07$) \\
Var002 & 21.968 & 0.849 & 0.934 & --0.07 & 5.52 \\
 & ($\pm 0.03$) & ($\pm 0.03$) & ($\pm 0.03$) & ($\pm 0.10$) & ($\pm 0.10$) \\
Var003 & 21.510 & 0.166 & --0.543 & 0.38 & 5.31 \\
 & ($\pm 0.03$) & ($\pm 0.05$) & ($\pm 0.04$) & ($\pm 0.08$) & ($\pm 0.08$) \\
Var004 & 23.052 & --0.305 & --0.758 & -- & -- \\
 & ($\pm 0.12$) & ($\pm 0.12$) & ($\pm 0.19$) & -- & -- \\
Var005 & 21.344 & --0.302 & --0.222 & -- & -- \\
 & ($\pm 0.01$) & ($\pm 0.01$) & ($\pm 0.03$) & -- & -- \\
Var006 & 21.767 & --0.145 & --0.017 & -- & -- \\
 & ($\pm 0.01$) & ($\pm 0.02$) & ($\pm 0.02$) & -- & -- \\
Var007 & 22.053 & 1.264 & 0.800 & 0.31 & 5.23 \\
 & ($\pm 0.04$) & ($\pm 0.03$) & ($\pm 0.03$) & ($\pm 0.13$) & ($\pm 0.13$) \\
Var008 & 21.634 & 0.523 & 0.094 & -- & -- \\
 & ($\pm 0.01$) & ($\pm 0.02$) & ($\pm 0.02$) & -- & -- \\
Var009 & 20.566 & --0.237 & --0.151 & 0.13 & 4.27 \\
 & ($\pm 0.01$) & ($\pm 0.01$) & ($\pm 0.01$) & ($\pm 0.08$) & ($\pm 0.08$) \\
Var010 & 22.466 & 0.566 & 0.264 & --0.29 & 5.52 \\
 & ($\pm 0.02$) & ($\pm 0.02$) & ($\pm 0.02$) & ($\pm 0.15$) & ($\pm 0.15$) \\
Var011 & 22.760 & 1.283 & 0.702 & 0.09 & 4.68 \\
 & ($\pm 0.03$) & ($\pm 0.03$) & ($\pm 0.03$) & ($\pm 0.25$) & ($\pm 0.25$) \\
Var012 & 23.240 & 0.230 & 0.279 & 0.38 & 5.82 \\
 & ($\pm 0.05$) & ($\pm 0.06$) & ($\pm 0.06$) & ($\pm 0.20$) & ($\pm 0.20$) \\
Var013 & 22.973 & 0.588 & 0.545 & -- & -- \\
 & ($\pm 0.02$) & ($\pm 0.03$) & ($\pm 0.03$) & -- & -- \\
Var014 & 23.240 & 0.590 & 0.345 & --0.15 & 5.15 \\
 & ($\sim 0.03$) & ($\pm 0.03$) & ($\pm 0.04$) & ($\pm 0.25$) & ($\pm 0.25$) \\
Var015 & 23.799 & 0.776 & 0.272 & --0.04 & 6.41 \\
 & ($\pm 0.05$) & ($\pm 0.05$) & ($\pm 0.05$) & ($\pm 0.18$) & ($\pm 0.18$) \\
\tableline
\end{tabular}
\end{center}
\caption{Multi-wavelength Measurements}
\tablenotetext{a}{Measured from December 23, 2003 MegaCam images.}
\tablenotetext{b}{Measured from mean of IRAC images recorded on March 18, 2014 and 
October 10, 2014.}
\end{table*}

\subsection{CMDs and IR Colors} 

	The locations of the variables on the $(i', g'-i')$ CMD are indicated in Figure 
5, where magnitudes and colors from the December 2003 MegaCam observations are shown. 
Areas of the CMD that are expected to contain Cepheids and LBVs are also marked. The 
Cepheid instability strip in Figure 5 is that specified by Di Criscienzo et al. (2013), 
and the variables identified here as Cepheids fall within its boundaries. 
The blue and red boundaries of the region that contains LBVs shown in 
Figure 1 of Smith et al. (2004) are also indicated in Figure 5. The area 
of the CMD that contains LBVs appears to extend to fainter magnitudes than the 
classical S Dor strip (Smith et al. 2019). The blue S Dor 
sequence coincides with a discontinuity in the NGC 247 CMD. 
The two most luminous LBV candidates found from the GMOS observations 
using the dispersion in the magnitude measurements have intrinsic 
colors and magnitudes that place them on the S Dor instability strip, although spectra 
will be required to assign these secure LBV designations.

	The expected brightness of the S Dor strip in Figure 5 is consistent 
with the brightnesses of LBVs in other galaxies that have similar 
M$_K$ as NGC 247, and hence likely have similar metallicities. M33 
is of particular interest as it has a well-populated intrinsically bright blue plume 
in its CMD, and a number of LBVs and candidate LBVs have been identified 
(e.g. Massey et al. 2007; Humphreys et al. 2017 and references therein). These have 
$V$ between 16 and 18 (Martin \& Humphreys 2017), which corresponds to 
$V \sim 19.5 - 21.5$ at the distance of NGC 247. Given that the $g'$ and $V$ magnitudes 
of blue stars differ by only a few tenths of a magnitude (Fukugita et al. 
1996) then the LBV population of M33 thus overlaps with the area in Figure 5 
that is expected to contain LBVs after accounting for the difference in distance.

	NGC 2403 is a late-type spiral galaxy in the M81 group with 
a distance modulus similar to that of NGC 247, and the LBVs V37 and V38 in that 
galaxy have $V = 20.6$ and 19.4, respectively (Humphreys et al. 2019). As with the 
LBVs in M33, the brightnesses of these objects overlap with the area marked 
in Figure 5 that is expected to contain LBVs. We emphasize that the region of 
the CMD that contains LBVs also contains other types of intrinsically bright blue 
variables (e.g. Martin \& Humphreys 2017), so a location on the CMD that is close to the S 
Dor instability strip is not an ironclad indicator of a classification as an LBV.
 
\begin{figure}
\figurenum{5}
\epsscale{0.9}
\plotone{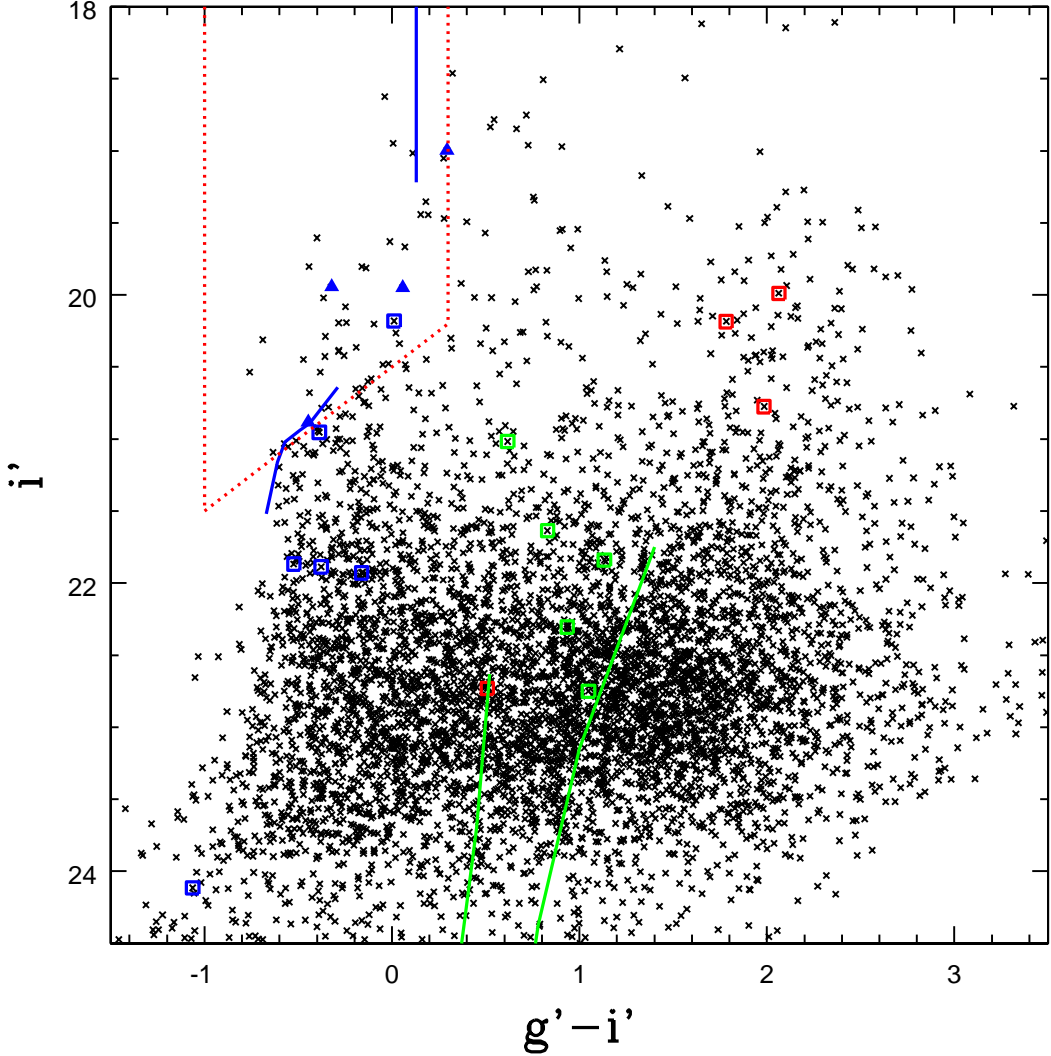}
\caption{$(i', g'-i')$ CMD of stars in the southern half of NGC 247 
constructed from the December 2003 MegaCam observations. The locations of all variables 
found in this paper are marked. Early-type variables (blue squares), Cepheids (green 
squares), and semi-regular variables (red squares) identified using the procedure 
described in Section 3 are indicated with open symbols. The four early-type variables 
identified using location in the CMD in Section 6 are shown as solid blue triangles.
The green lines mark the boundaries of the full Cepheid instability 
strip from Di Criscienzo et al. (2013). The blue lines mark the region 
occupied by LBVs in Figure 1 of Smith et al. (2004). The LBV boundaries 
were translated onto the observational plane using relations from de Jager 
\& Nieuwenhuijzen (1987), Fukugita et al. (1996), and Allen (2000). 
The red dashed lines indicate the region of the CMD used in Section 6 to isolate 
possible LBVs. The placement of the fiducial sequences assume a distance modulus of 
27.64 and E(B-V) = 0.18 (Gieren et al. 2009).}
\end{figure}
 
	Because they encompass a diverse group of objects, semi-regular variables 
are found over a range of locations in the CMD. Var012 falls near the blue edge of the 
Cepheid instability strip. As discussed previously, a secure period could not be 
identified for this star, leading us to assign it class SRd. As for the 
other semi-regular variables, these fall at or near the bright end of the red plume 
in the CMD, as expected if they are heavily evolved supergiants. The absence of 
periodicity suggests a classification of type SRb for these stars. At least some of 
these stars have well known analogs in the Galaxy. The classical semi-regular 
variable star Betegeuse ($\alpha$ Ori) has M$_{i'} \sim -8$, which would 
correspond to $i' \sim 20$ in NGC 247. This is comparable to the brightest 
semi-regular red variables detected here.

	Davidge (2021) discussed the ([3.6], [3.6]--[4.5]) CMD of NGC 247. 
The peak of the stellar sequence in NGC 247 was found to occur near 
[3.6] = 17 based on a comparison of number counts in NGC 247 with those in a 
control field that is offset from the galaxy. Intrinsically rare 
sources in NGC 247 might then be brighter than [3.6] = 17, given that they may not 
occur in numbers that would elevate them in a statistically significant manner 
above those of foreground and background objects. 
In fact, the entries in Table 4 indicate that all three of the candidate 
LBVs that are detected by SPITZER are brighter than [3.6]=17, reflecting 
the highly evolved nature and intrinsic luminosity of these stars. 
In contrast, all of the Cepheids detected in the SPITZER observations have 
[3.6] magnitudes that are comparable to or below the brightness limit 
estimated by Davidge (2021) -- the brightest of the Cepheids detected by SPITZER is 
Var010, for which [3.6] = 16.95.

	[3.6]--[4.5] and $g'-$[3.6] colors are listed in the last two columns of 
Table 4. [3.6]--[4.5] colors that exceed those produced by stellar photospheres 
are a signature of hot dust emission, and none of the 
variables have [3.6]--[4.5] colors that depart significantly from the 
scatter envelope of the NGC 247 stellar sequence in Figure 6 of Davidge (2021). 
With the caveat that the SPITZER and MegaCam observations were not recorded at the 
same epoch, it is also apparent that the $g'-[3.6]$ colors of the candidate LBVs are not 
markedly redder than those of the other stars in the sample. Additional evidence for the 
pedestrian infrared photometric properties of the candidate LBVs in 
NGC 247 comes in the form of the mixed success of detection with SPITZER: three 
of the candidate LBVs were not detected in the SPITZER images. The 
photospheric-like IR colors of these stars does not mean that 
hot dust is abscent. Rather, if hot dust is present then the 
emission is smaller than that produced by the stellar photospheres at these wavelengths. 
While there is no evidence for an IR excess in the candidate semi-regular variables and 
LBVs in the SPITZER IRAC observations, the possibility of a cool dust shell around 
these objects can not be discounted without observations at longer wavelengths.

\subsection{Location in NGC 247}

	The locations of the variables discovered in this paper are 
shown in Figure 6. The background is the median 1 arcsec 
FWHM GMOS image that served as the reference for the photometric 
measurements. The locations of Cepheids are shown in the right hand 
panel, while those of all other variables are indicated in the left hand panel.

\begin{figure}
\figurenum{6}
\epsscale{1.0}
\plotone{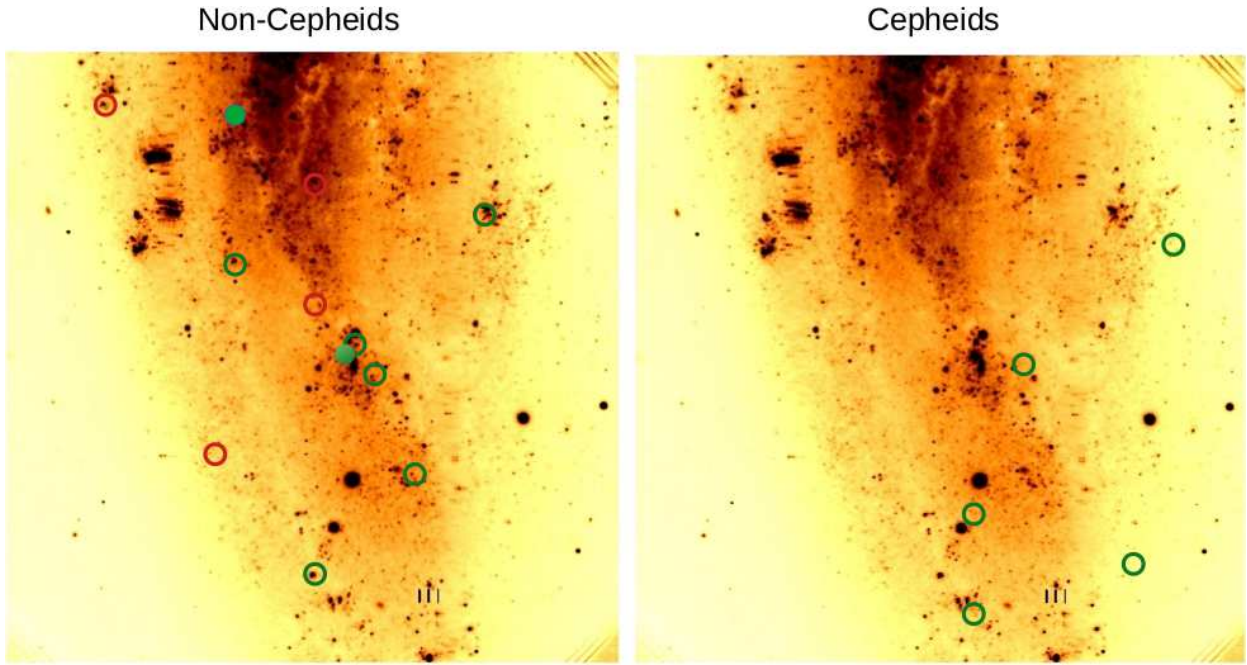}
\caption{Locations of Cepheids (right hand panel) and other variables (left hand panel) 
in the median 1 arcsec FWHM $g'$ image. Semi-regular variables and candidate LBVs 
identified using the procedures described in Section 3 are 
indicated with red and green circles. The locations of the two cLBVs identified using 
the procedure described in Section 6 are indicated with filled green circles. The Cepheids 
and non-Cepheids have different locations in the sense that Cepheids tend to avoid 
star-forming knots, and this is consistent with ages of at least a few tens of Myr. 
In contrast, many of the semi-regular variables and cLBVs are in star-forming 
complexes, as might be expected if they are young massive evolved stars.}
\end{figure}

	Cepheids and non-Cepheids tend to be located in different parts of 
NGC 247. The majority of blue and semi-regular variables tend to be 
in or near collections of stars that are likely star-forming complexes and/or 
associations. This is consistent with these objects being young massive stars.
In fact, Rodriguez et al. (2019) found that many of the most luminous blue stars in this 
part of NGC 247 fall along a serpentine structure 
that is associated with the edges of large-scale bubbles (Davidge 2021). 
The majority of the non-Cepheid variables fall along or are close to this 
serpentine structure. 

	The Cepheids tend to be found away from star-forming 
knots and associations. This is consistent with these objects having 
ages in excess of tens of Myr, thereby giving them time to diffuse 
away from their places of birth, and/or for their birth 
places to have dissolved. There is a selection effect when identifying Cepheids, as 
shorter period Cepheids as well as those with smaller amplitudes will be 
more susceptible to blending in dense stellar regions. This being said, shorter 
period Cepheids have older ages than their brighter, longer period 
cousins. Therefore,they will have had more time to dissociate from their natal environment 
and migrate to comparatively low density regions. 

\section{LIGHT CURVES OF PHOTOMETRICALLY SELECTED LUMINOUS BLUE STARS}

	The $5\sigma$ dispersion criterion for selecting variables that was discussed in 
Section 3 relies on brightness measurements made from the ten images that have $\sim 1$ 
arcsec FWHM angular resolution. While this is a robust means of detecting objects 
with large amplitude light variations during the course of the observing campaign, it 
sets a high standard for selecting variables. However, a high standard is necessary to 
suppress specious detections when large samples of stars are examined. 

	Another approach is to target sparsely populated 
locations in the CMD where a high frequency of variables 
might be expected, as a relaxed criterion for variability can then be applied. 
Such a procedure might also facilitate the identification of 
non-periodic and low amplitude eruptive variables -- if 
large scale variability events did not occur when the images with the best image 
quality were recorded, then these stars may avoid detection when applying 
the $5\sigma$ dispersion threshold discussed in Section 3. It might also be possible 
to identify subtle long term trends and low amplitude systematic trends in light curves 
that could evade detection in the dispersion measurements.

	The brightest blue stars are obvious targets for pre-selection based on location 
in the CMD. Many highly evolved massive stars are variable. In addition, the expected 
contamination from foreground objects will also be modest if one focuses on blue colors. 
Proximity to the location of star-forming regions provides an additional means 
of restricting contamination from sources not associated with the target galaxy, 
while obvious background galaxies with blue colors can also be identified 
based on their appearance. Finally, because they are 
among the brightest objects in a galaxy, luminous blue stars also have 
photometeric measurements with comparatively high S/N ratios -- low amplitude 
variations in light might be found that would go undetected in other, fainter stars 
in the same galaxy. 

\subsection{Photometry of a Known Bright Blue Variable in NGC 247}

	Before applying this approach to search for variability in the brightest blue 
stars, we first examine the photometric properties of a previously studied luminous star 
in NGC 247. Solovyeva et al. (2020) used spectra and photometry to identify a candidate 
LBV in NGC 247 (J004702.18--204739.9), as well as a B$[e]$ supergiant 
(J004703.27--204708.4). The light curve of the 
former shows evidence of episodic variations with a $\sim 1$ magnitude amplitude
in Figure 4 of Solovyeva et al. (2020), whereas that of the latter 
varies by $\sim 0.2 - 0.3$ mag in $V$. Unfortunately, 
J004702.18--204739.9 falls in one of the gaps between the GMOS CCDs and so 
reliable photometry of that star can not be extracted from the GMOS images. 
However, reliable photometry can be obtained for J004703.27--204708.4. 
The GMOS photometry is of interest for that star 
as it samples a portion of the light curve not 
covered previously. Perhaps of greater relevance for the present study is that 
the sampling cadence of the GMOS images is as small as 1 - 2 days, whereas the majority 
of photometric measurements for J004703.27--204708.4 summarized in Table 2 of 
Solovyeva et al. (2020) have a typical cadence of months or years. Does the 
GMOS photometry of this star show signs of variability?

	The light curve of J004703.27--204708.4 extracted from the GMOS observations 
is shown in Figure 7. The $g'$ magnitudes obtained from GMOS are 
a few tenths of a magnitude brighter than might be 
expected based on the $V$ magnitudes presented in Table 2 of Solovyeva 
et al. (2020). This could occur if the star was at a persistently bright 
phase of its light variations in mid to late 2008. The overall 
light level in the GMOS photometry may also be boosted by blending in the area immediately 
surrounding the star. Figure 6 of Solovyeva et al. (2020) 
shows that there are at least three moderately bright stars 
within a few tenths of an arcsec of the dominant source. To reduce possible contamination 
from nearby stars, Solovyeva et al. (2020) restricted their photometric measurements 
to images with an angular resolution $< 0.8$ arcsec, whereas those 
used to construct Figure 7 have an image quality $> 1$ arcsec.

\begin{figure}
\figurenum{7}
\epsscale{1.0}
\plotone{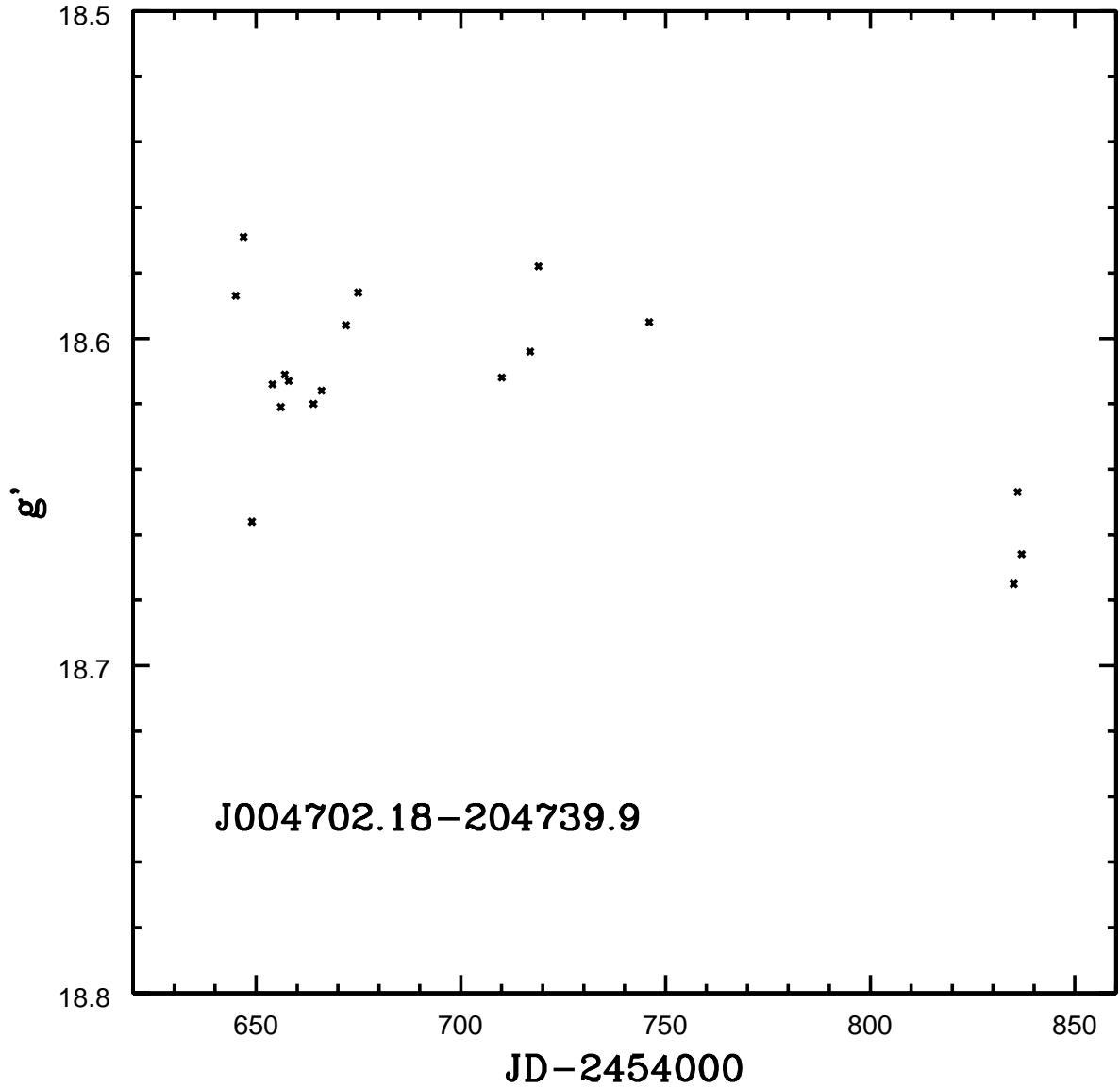}
\caption{The $g'$ light curve of J004702.18--204739.9, which Solovyeva et al. (2020) 
classify as a B$[e]$ supergiant. Only measurements obtained from 
the GMOS images are shown. There is a $\sim 0.1$ magnitude throw in the GMOS measurements, 
and a hint of a long term decrease in mean brightness from mid 2008 to early 2009.}
\end{figure}

	There is jitter in the light curve in Figure 7, with an overall amplitude 
of $\sim 0.1$ magnitudes. There is also a tendency for J004703.27--204708.4 to become 
fainter towards later epochs. While the jitter is smaller in amplitude than 
that in the Solovyeva et al. light curve, the trend in declining brightness 
suggests that the variations seen in their light curve may occur on time scales of a 
month or more. The 1$\sigma$ scatter of other stars with this 
brightness that were not flagged as variables is $\pm 0.02 - 0.03$ mag, which is 
smaller than the scatter in Figure 7. J004702.18--204739.9 was not flagged as 
a variable star in Section 3 because the scatter does not meet the 5$\sigma$ 
selection criterion used there. Still, signs of variability are present in Figure 7.

\subsection{New Bright Blue Variables in NGC 247}

	Given the likelihood that as yet undiscovered variables lurk in the 
photometric measurements, we have examined the light curves of objects within the red 
dashed lines in Figure 5. After limiting the sample to those sources that 
fall within the main body of the galaxy to reduce contamination from 
foreground/background objects, and rejecting obvious background 
galaxies, we find that the light curves of most objects show 
only marginal scatter in the GMOS observations: if variability 
occurs in these objects then it must be on timescales that exceed the six months 
that was sampled with GMOS. 

	After rejecting objects that have non-stellar morphologies, and hence are 
unresolved blends, there are two sources that show obvious scatter in the GMOS 
observations, indicating changes in light levels over the course of a few days 
to a few months. The light curves of these stars are shown in Figure 8, where 
departures from mean light levels on a scale of many tenths of 
a magnitude are seen. Neither light curve shows signs of periodic behaviour. 
These objects would have been flagged as variables if a lower than $5\sigma$ 
dispersion criterion had been applied in Section 3.

\begin{figure}[!ht]
\figurenum{8}
\epsscale{1.0}
\plotone{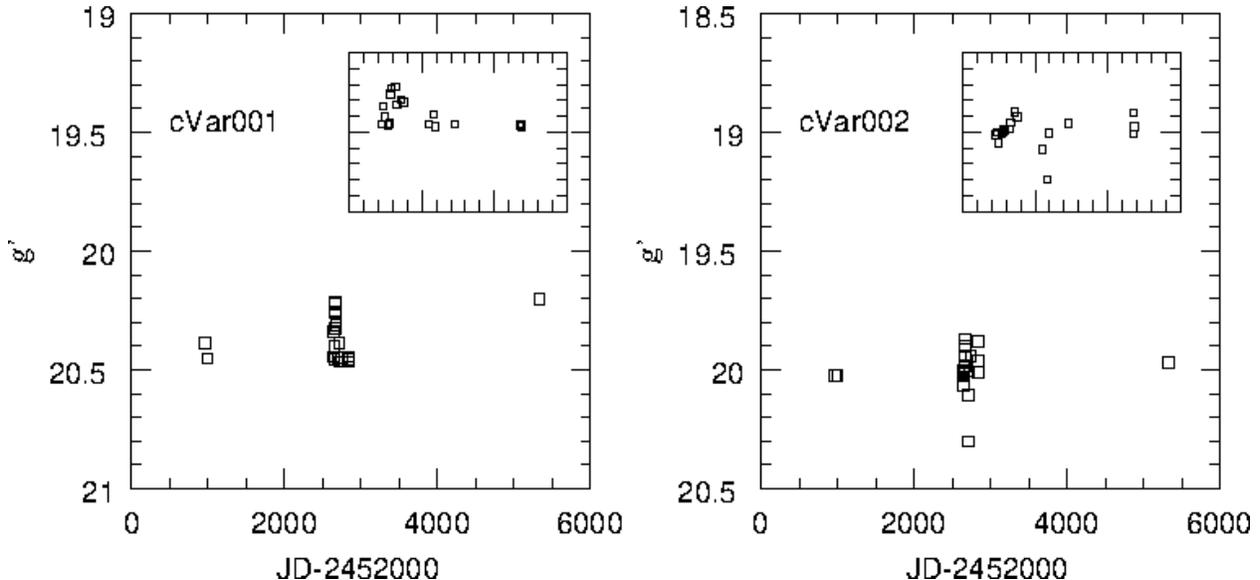}
\caption{Light curves of luminous early-type variable stars in the disk of NGC 247 that 
were selected based on location in the CMD and are not obvious blends or background 
galaxies. The inset in each panel shows the GMOS data between JDs 2454600 and 2454900, 
with a 1 magnitude throw along the vertical axis.}
\end{figure}

	The locations and photometric properties of these objects are listed in Table 5, 
while finding charts that cover $15 \times 15$ arcsec are shown in Figure 9. cVar001 is 
separated by $\sim 2$ arcsec from a star-forming complex, and it is possible 
that some of the variations in the light curve of this star may be affected by 
contamination from brighter nearby objects, especially in the images in the 1.5 arcsec 
FWHM grouping. Therefore, we consider the cVar001 light curve to be tentative. 
In contrast, cVar002 is relatively isolated, and so less prone to seeing-related 
variations in the light curve.

\begin{table*}[!ht]
\begin{center}
\begin{tabular}{cccccccc}
\tableline\tableline
Name & RA & Dec & g' & g'--r' & r'--i' & [3.6] & [3.6]--[4.5] \\
 & (2000) & (2000) & & & & & \\
\tableline
cVar001 & 00:47:10.3 & --20:48:53.9 & 20.437 & -0.031 & -0.416 & -- & -- \\
cVar002 & 00:47:06.1 & --20:46:45.1 & 20.007 & 0.041 & 0.016 & 17.23 & 0.09 \\
\tableline
\end{tabular}
\end{center}
\caption{Blue Variables Identified from CMD Selection}
\end{table*}

\begin{figure}[!ht]
\figurenum{9}
\epsscale{1.0}
\plotone{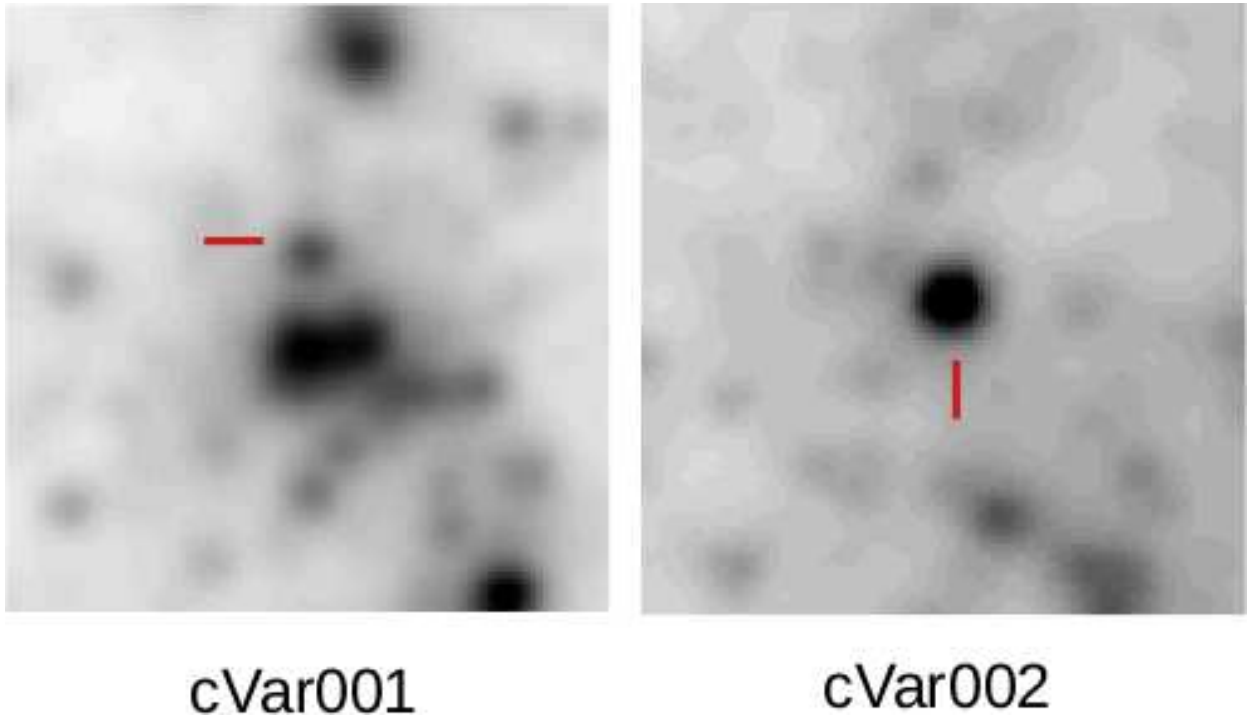}
\caption{Same as Figure 1, but showing finding charts for the variables 
identified using a location on the CMD that corresponds to that of 
LBVs in the Galaxy and the Magellanic Clouds. Note that only 
cVar002 is isolated, as cVar001 is part of a star-forming complex.}
\end{figure}

	We have attempted to identify these objects in the SPITZER images examined by 
Davidge (2021). The SPITZER images have a poorer angular 
resolution than the GMOS observations, and cVar001 and the star-forming 
complex immediately to the south of it appear as a single object in the SPITZER data. 
Therefore, [3.6] and [4.5] measurements for this star are not given in Table 5. 
As for cVar002, it has [3.6]--[4.5] colors that do not depart from the primary 
sequence of objects in the $([3.6], [3.6]-[4.5]$ CMD of NGC 247 
shown in Davidge (2021), indicating that if emission from 
hot dust is present then it does not dominate over the photospheric light from the 
host object.

\section{SUMMARY \& DISCUSSION}

	Deep $g'$ images obtained with GMOS on GS have been used to search for variable 
stars in the southern half of the nearby disk galaxy NGC 247. This part of NGC 247 
contains areas of on-going and recent (i.e. within a few hundred 
Myr) star formation (Rodriguez et al. 2019; Davidge 2021), and so can be expected to 
harbor a rich population of luminous variable stars that span a range of types. The GMOS 
observations cover a 7 month time period, with intensive coverage over a one month 
period. These observations are thus sensitive to periodic changes in light levels 
on time scales of a few days to tens of days. A mix of periodic and non-periodic 
variables are identified. 

	The time coverage has been extended using archival observations from the CFHT 
MegaCam and NOIRLab DECam. An extended time baseline is of greatest interest 
for variables that may have erratic light variations, such as LBVs. Observations from the 
SPITZER IRAC are also examined to search for evidence of hot circumstellar dust. 
A summary of the main findings is as follows:

\vspace{0.3cm}
\noindent{\bf 1)} Of the fifteen new variables that have been discovered, three are 
Cepheids. These were identified based on the dispersion in the magnitude measurements 
made from the ten GMOS images that have $\sim 1$ arcsec FWHM image quality. The newly 
discovered Cepheids have periods that are near the low end of those already discovered in 
NGC 247. Two previously identified Cepheids were also recovered. 

	Four Cepheids that were discovered by Garcia-Valera et al. (2008) in the area 
observed with GMOS were not detected. One failed detection was because the Cepheid 
(cep013) is located close to the edge of the GMOS science field, and 
an inspection of the GMOS photometry of that object reveals light 
variations that are consistent with those of a Cepheid. The failure to detect the other 
Cepheids is likely due to the shorter time coverage of the GMOS data when compared 
with that employed by Garcia-Valera et al. (2008). A shorter time baseline 
makes the sampling of variability susceptible to stochastic effects. 

	In addition to serving as standard candles, Cepheids probe 
the SFH of the host galaxy during the past few hundred Myr. The progenitors of 
fundamental mode Cepheids with periods like those found in this study 
likely formed within the past 100 Myr (e.g. Bono et al. 2005). This is shorter than the 
$\sim 0.6$ Gyr disk crossing time that results for NGC 247 
if a rotational velocity of $\sim 100$ km/sec at the optical 
diameter of 19.9 arc minutes (Carignan \& Puche 1990) and a 
distance of 3 Mpc are adopted. Hence, Cepheids with periods like those found to date 
likely have not had enough time to mix throughout the NGC 247 disk. Still, Cepheids 
will acquire random motions from interactions with molecular 
clouds, and they might have attained random velocities of $\sim 20 - 
30$ km/sec if they have reached the equilibrium velocity dispersion of a host disk with 
a 100 km/sec rotation velocity (e.g. Kregel et al. 2005). 
Cepheids in the NGC 247 disk might then be expected to move projected 
distances of only 2 -- 3 kpc from their places of birth over 100 Myr. The present 
study has found Cepheids in the southern disk that have periods that are comparable to 
the shortest period Cepheids found in the northern disk. That Cepheids with 20 -- 30 
day periods are seen over much of the NGC 247 disk overcomes 
at least in part the lop-sided distribution of the Cepheids 
found by Garcia-Valera et al. (2008).

\vspace{0.3cm}
\noindent{\bf 2)} Eight new variables with blue colors have been discovered. 
Six of these were identified solely on the dispersion in their magnitude 
measurements with no knowledge of their colors, while two more were identified by 
targeting the area of the CMD that contains LBVs and then applying relaxed 
conditions to identify variability. The two brightest variables 
identified at the time the MegaCam data were recorded 
using the first technique (Var001 and Var009) have photometric properties that place 
them close to the S Dor instability strip in the $(i', g'-i')$ CMD. 
Depending on the epoch of observation and the amplitude of light variations, 
some of the fainter blue variables discovered using this technique 
may prove to have brightnesses and colors that could position them in 
the LBV region of the CMD.

	Two differing models have been proposed to explain the nature of LBVs. 
Humphreys \& Davidson (1994) review the photometric characteristics of 
LBVs, and suggest that they are massive single stars, and can be sorted 
into two group. The first group consists of the most massive stars that do 
not evolve to the RSG stage due to large-scale mass loss. These stars show large 
amplitude photometric variations that may be tied to mass 
ejection. The second group is fainter than the first, and shows less dramatic 
light curve variations. The stars in that group are thought to be experiencing mass loss 
after evolution as a RSG.

	Smith \& Tombleson (2015) and Smith (2017) present an alternate interpretation, 
suggesting that LBVs are massive blue stragglers that are the 
product of binary evolution. This is based on the finding that LBVs 
appear to be more isolated than the Wolf-Rayet stars that are thought to be the end point 
of massive single star evolution (but see Humphreys et al. 2016). 
In addition, there is evidence for supernovae that 
appear to have originated from massive stars that did not shed large quantities of their 
mass prior to explosion, which challenges the classical notion of 
mass loss throttling the evolution of massive stars. 
While Figure 6 indicates that there is an association between 
the cLBVs and areas of star formation in NGC 247, Figures 1 and 9 
suggest that the majority of the early-type variables found here are embedded in 
complex environments, in the sense that they have one or more nearby companions, 
although cVar002 is an exception.

	Wofford et al. (2020) review the statistics of LBVs and cLBVs 
in nearby galaxies, and emphasize the importance of identifying LBVs in a range of 
galaxy environments. NGC 247 is a potentially important target in 
this regard as it is one of only four galaxies searched so far for LBVs 
at visible wavelengths that show obvious tidal features. Table 1 of 
Wofford et al. (2020) lists three galaxies that show evidence of tidal 
distortion at the present day: the SMC, NGC 55, and DDO 68. These galaxies have 
M$_K$s that are 1 -- 2 magnitudes fainter than NGC 247, and together they contain 3 LBVs 
and 6 cLBVs. The present survey of the southern half of 
NGC 247 has found a comparable number of intrinsically bright blue variables.

	The presence of tidal features might affect the 
incidence and spatial distribution of LBVs and related objects 
within a galaxy. As more galaxies with tidal distortions are searched for LBVs then 
this might provide insights into -- say -- the role that tidally-induced turbulence 
in the ISM plays in defining the frequency of LBVs in a galaxy. Indeed, environments that 
foster the formation of the largest giant molecular clouds will likely also have the 
highest probability of forming the most massive stars (e.g. Jerabkova et al. 2018). 
In the context of the model in which LBVs are single massive stars then one might 
expect a correlation between the incidence of LBVs and the masses of molecular clouds.
Finally, galaxies that show prominent signs of tidal interaction 
also tend to have high SFRs concentrated in their central regions 
(e.g. Pan et al. 2019), and so the specific frequency of LBVs might be higher there.

	M82 is a nearby tidally distorted galaxy that is a prime 
target to search for LBVs given its high SFR, although 
the extinction towards the active star forming regions make studies at visible 
wavelengths problematic. Neverthess, using radio observations as an indicator, 
Matilla et al. (2013) identify an object with 
an absolute magnitude consistent with that of an LBV in outburst, although the nature 
of this source remains enigmatic. Surveys at longer wavelengths may also have found 
cLBVs in a range of other galaxies. The SPIRITS survey (Kasliwal et al. 2017) has 
identifed IR transients that are associated with star-forming galaxies. These so-called 
SPRITES have photometric properties that are reminiscent of LBVs in eruptive phases.

	LBVs are very rare objects, and are but one type of variable identified among 
intrinsically bright blue stars. Thus, while we have 
identified candidate LBVs in this study, we anticipate that 
subsequent observations will reveal that many -- if not all -- of the 
objects so identified will be found not to be true LBVs. This being said, 
{\it bona fide} LBVs may be missed in the present survey given that these objects 
can experience extended periods of photometric dormancy (Humphreys \& Davidson 1994).

	The 1 arcsec FWHM angular resolution of the GMOS images coupled with the 
crowded nature of star forming regions also means that many of the blue variables may 
be blended with stars of comparable (i.e. within 1 -- 2 
magnitudes) brightness. An inspection of HST images taken as 
part of the ANGST survey (Dalcanton at el. 2009) reveal that Var001 is actually a tight 
knot of stars, although there is a dominant star in this asterism that is $\sim 2$ 
magnitudes brighter than its companions. In contrast, Var009 is an isolated object. 

	Establishing the nature of the brightest blue variables found here will require 
spectroscopy and additional photometry to characterize the absorption and emission 
lines in their spectra, and to better characterize the time scale and amplitude of light 
variations. A possible hint as to the nature of the intrinsically bright 
blue variables found here is that none appear to show excess emission in 
the $3 - 5\mu$m wavelength region. Despite their location in the 
CMD close to the S Doradus instability strip, there is thus no evidence for 
hot circumstellar dust shells, as might form from metal-enriched material ejected 
throughout their evolution. Of course, dusty envelopes that do not 
dominate the emission in the $3 - 5\mu$m wavelength region
might still be present. A search for cooler dust emission at 
longer wavelengths where the contribution from photospheric light is 
suppressed awaits measurements with arcsec or better angular resolutions.

\vspace{0.3cm}
\noindent{\bf 3)} The original motivation for the GMOS observing program that serves as 
the basis for this paper was to search for EBs that contain massive stars 
on or near the main sequence. Assuming that massive main 
sequence stars have M$_V \sim -5$ to --6 then such EBs will have $g' \sim 
22 - 23$ in NGC 247. While some of the blue variables found here have brightnesses in 
this range, their light curves do not show variations that are consistent with EBs. 
However, one of the blue variables that falls well below the S Dor instability strip 
on the $(i', g'-i')$ CMD shows periodic variations that are sinusoidal 
in nature. We speculate that these variations are the result of 
star-to-star interactions in a close binary system, with 
the variations presumably related to spot activity on one of the stars. If the period 
found here is off by a factor of two then the light curve would be that of an ellipsoidal 
variable. Radial velocity measurements of this object that cover many tens of 
days should reveal if it is part of a binary system and -- if so -- the true period 
of the variability.

\vspace{0.3cm}
\noindent{\bf 4)} Two cLBVs were found by targeting the region of the CMD that contains 
the intrinsically brightest LBVs (see above). The frequency of variability among stars 
in this part of the CMD may provide clues into the nature of these objects and 
the chances of their detection in surveys that are similar to this one. 
In fact, the time series photometric properties of the majority 
of intrinsically luminous blue stars in the main body of the galaxy 
showed little or no signs of variability from June 2008 to January 2009. 
The frequency of detection over this time frame was $\frac{2}{10} = 0.2$, with 
a threshold amplitude for variability detection $\sim 0.1$ magnitudes. 
The photometrically stable stars in this part of the CMD may simply have been 
observed during a quiet phase of their evolution. 
It is unlikely that the photometrically stable stars are in the foreground 
given that they are seen against in the main body of NGC 247 and are close 
to star forming regions. Obvious background galaxies were also deleted from the sample. 

\vspace{0.3cm}
\noindent{\bf 5)} The photometric variablity of the spectroscopically classified 
B$[e]$ star J004702.18--204739.9, which is one of the brightest stars in NGC 247, has 
been examined with the GMOS images. These data have a sampling cadence that is 
finer than measurements used by Solovyeva et al. (2020) to 
examine the photometric stability of this star. Non-periodic 
variations of a few hundredths of a magnitude over timescales of a few days are found, 
with a $\sim 0.1$ magnitude systematic decrease in $g'$ over a six month time span. 
J004702.18--204739.9 is in a crowded environment and is likely blended with other 
objects in the GMOS images. Thus, the intrinsic variations in the dominant star 
in the blended conglomeration are probably larger than those seen here.

\vspace{0.3cm}
\noindent{\bf 6)} Four semi-regular variables have also been identified. Three of 
these have red colors, and are classified as type SRb. These are located in 
or near dense stellar knots and associations, which is consistent with 
them being young, highly evolved objects. These stars have intrinsic 
brightnesses that are similar to those of the classic Galactic semi-regular 
variable Betelgeuse ($\alpha$ Ori). 

	The fourth semi-regular variable has a g'--i' color that is indicative 
of an intermediate spectral-type, and a light curve that shows variations 
that are reminiscent of those associated with RCrB stars. It has been classified 
as type SRd in this study. Longer term photometric monitoring of this star will be 
useful to determine if it shows the persistent moderately long-term deep dips in its 
brightness that are the defining photometric characteristic of RCrB stars. RCrB 
stars are rare, and may be the result of a merger of white dwarfs (e.g. Tisserand et al. 
2020), making them of inherent interest for probing the most advanced and exotic stages 
of binary star evolution. To the best of our knowledge, this is the first candidate 
RCrB star to be identified outside of the Local Group. A survey for stars 
of this type among nearby galaxies will provide statistics that will help identify the 
nature of these objects.

\vspace{0.3cm}

	Surveys for variable stars in NGC 247 have so far 
explored only the tip of the iceberg, and there is much room for future exploration. 
The detection of variable stars with smaller amplitudes than those found here 
and/or of fainter sources will require deeper time series photometry with sub-arcsecond 
angular resolutions. Such observations would enable a search for fainter Cepheids with 
shorter periods than those found to date in NGC 247. 
The Cepheids found in such a survey could be used to examine the spatial extent 
of star formation in NGC 247 a few hundred Myr in the past, which is an epoch when 
there is evidence for large-scale star formation in at least some parts of the galaxy. 
High angular resolution observations will also enable the 
exploration of intrinsically crowded environments, such as the dense star forming 
complexes where LBVs might be present.

	While massive stars in binary systems are likely the norm (e.g. Sana et al. 2012), 
the detection of massive EBs in NGC 247 has so far remained elusive. However, 
the detection and characterization of even one or two of these will provide a fundamental 
constraint on the distance to NGC 247 that is based on purely geometric arguements. 
While challenging, obtaining the precise radial velocity curves for massive EBs in 
NGC 247 and other galaxies with similar distances that are required for such a 
project should be possible with the up-coming generation of 30 - 40 meter telescopes.

\acknowledgements{It is a pleasure to thank the anonymous referee for a timely 
report and thoughful comments that greatly improved the manuscript.}

\end{document}